\documentclass[aip,reprint]{revtex4-1}

\usepackage[fleqn]{amsmath}
\usepackage{amssymb}
\usepackage{dcolumn}
\usepackage{graphicx}
\usepackage{subfigure}
\usepackage{nccmath}
\usepackage{bm}

\newcommand{\edotk}[1]{\bm{e}^*_T\cdot\bm{\kappa}_{#1}}

\draft 

\begin{document}


\title{First-order thermal correction to the quadratic response tensor and rate for second harmonic plasma emission} 

\author{B. Layden}
\email{b.layden@physics.usyd.edu.au}
\affiliation{School of Physics, University of Sydney, Sydney, N.S.W. 2006, Australia}
\author{D. J. Percival}
\affiliation{Defence Science and Technology Organisation, P.O. Box 1500, Edinburgh, S.A. 5111, Australia}
\author{Iver H. Cairns}
\affiliation{School of Physics, University of Sydney, Sydney, N.S.W. 2006, Australia}
\author{P. A. Robinson}
\affiliation{School of Physics, University of Sydney, Sydney, N.S.W. 2006, Australia}

\date{\today}

\begin{abstract}

Three-wave interactions in plasmas are described, in the framework of kinetic theory, by the quadratic response tensor (QRT). The cold-plasma QRT is a common approximation for interactions between three fast waves. Here, the first-order thermal correction (FOTC) to the cold-plasma QRT is derived for interactions between three fast waves in a warm unmagnetized collisionless plasma, whose particles have an arbitrary isotropic distribution function. The FOTC to the cold-plasma QRT is shown to depend on the second moment of the distribution function, the phase speeds of the waves, and the interaction geometry. Previous calculations of the rate for second harmonic plasma emission (via Langmuir-wave coalescence) assume the cold-plasma QRT. The FOTC to the cold-plasma QRT is used here to calculate the FOTC to the second harmonic emission rate, and its importance is assessed in various physical situations. The FOTC significantly increases the rate when the ratio of the Langmuir phase speed to the electron thermal speed is less than about 3.

\end{abstract}




\maketitle

\section{Introduction}


Plasma emission, which is the generation of radiation at multiples of the local electron plasma frequency $f_p$, is observed in various locations and phenomena in our solar system; these include type II \cite{wild50,wild63,cane82} and type III \cite{wild50,wild63,suzuki85,robinson00} solar radio bursts in the solar corona and interplanetary medium, and terrestrial foreshock emission.\cite{gurnett75,hoang81} Although several mechanisms have been proposed for these emissions, such as linear mode conversion \cite{field56,melrose80a,yin98} and cyclotron maser emission,\cite{wu02} they are generally attributed to three-wave interactions between Langmuir, transverse, and ion sound waves.\cite{ginzburg58,wild63,melrose80c,melrose85,cairns85}
Three-wave interactions include the coalescence of two waves to give a product wave, and the decay of one wave into two product waves. These processes occur due to the nonlinear response of the plasma medium to the wave fields. In kinetic theory, the response of a plasma to an electromagnetic disturbance is described by linear and nonlinear response tensors.\cite{tsytovich70,sitenko82,melrose80b,sagdeev69} Assuming the plasma response to be weakly nonlinear, induced plasma properties, such as the induced charge and current densities, can be expanded in powers of the amplitude of the electromagnetic field: this is termed the ``weak-turbulence expansion''. \cite{tsytovich77,melrose80b,sitenko82} The quadratic response tensor (QRT), defined as the coefficient of the second-order term in this expansion, describes the response of the plasma to two fields beating simultaneously to produce a third wave; therefore, the QRT is the relevant response tensor for three-wave interactions.



The general form of the QRT involves integrals over the velocity distribution function of the plasma particles. These integrals are difficult to evaluate exactly, and so the integrand is often approximated before performing the integrations, where the approximation that is made depends on the dispersion relations of the three wave modes involved. For interactions between three ``fast'' waves, that is, waves with a phase speed much greater than the thermal speed, the ``cold-plasma approximation'' to the QRT is often made, in which thermal effects are neglected in the description of wave coupling.\cite{melrose80b,sitenko82} This approximation lends itself to a simple derivation of the response tensor, and the resulting expression can be readily used in practical applications. However, the cold-plasma approximation becomes less accurate as the phase speed of one or more of the interacting waves approaches the thermal speed, and its range of validity is poorly defined.

Conversely, the QRT has been evaluated exactly for a thermal plasma by \citet{percival98a}, in terms of generalized plasma dispersion 
functions.\cite{percival98b} The resulting exact QRT accurately describes wave coupling in three-wave interactions between any wave modes. 
Although the use of the exact QRT is desirable, the expression is too cumbersome to apply analytically, and even its numerical evaluation presents 
difficulties as described in Sec.~II.

Due to the inaccuracy of the cold-plasma QRT at low phase speeds and the difficulty of applying the exact QRT, accurate approximations to the QRT are needed for a proper treatment of three-wave interactions. To this end, we derive here the first-order thermal correction (FOTC) to the cold-plasma QRT for a warm collisionless unmagnetized plasma, whose constituent particles have an arbitrary isotropic velocity distribution. The approximate response tensor derived in this paper, which is the sum of the cold-plasma QRT and its FOTC, has the advantages that it is more accurate than the cold-plasma QRT and more tractable than the exact QRT. This approximation is valid for interactions between three fast waves, such as the Langmuir-wave coalescence process that generates second harmonic emission. We find that the FOTC depends on the second moment of the distribution function, the phase speeds of the waves, and the interaction geometry.

First-order thermal corrections to the quadratic response have been derived before in the literature, but the cases treated are not suitable for modeling the Langmuir-wave coalescence process in space plasmas. For example, a FOTC to the cold-plasma longitudinal quadratic susceptibility was previously derived by \citet{sitenko82}; however, this quantity only describes interactions between three fast electrostatic waves and is thus inadequate for treating second harmonic plasma emission, in which electromagnetic transverse waves are produced. The FOTC to the cold-plasma QRT was derived in Ref.~\onlinecite{percival_thesis} for a Maxwellian velocity distribution of the plasma particles, but space plasmas are commonly observed to have power-law tails which must be modeled by a non-Maxwellian distribution, often the generalized Lorentzian (or ``kappa'') distribution.\cite{vasyliunas68,maksimovic97}  The expression that we derive is equivalent to that in Ref.~\onlinecite{percival_thesis} when the velocity distribution is Maxwellian, but allows the treatment of three-wave processes in non-Maxwellian plasmas. 

The rate of a three-wave interaction is dependent on the strength of the coupling between the waves, which is described by the QRT. Until now, the rate of second harmonic plasma emission via Langmuir-wave coalescence has been calculated assuming the cold-plasma approximation.\cite{cairns87a,willes96,li05} We use the FOTC to the cold-plasma QRT to calculate the FOTC to the rate of Langmuir-wave coalescence, and assess its contribution to the total interaction rate in various situations. There is a significant increase of the rate when the ratio of the Langmuir phase speed to the electron thermal speed is less than about 3.

The paper is structured as follows. In Sec.~II, the theory of response tensors and their derivation is described. In Sec.~III, we derive and discuss the FOTC to the cold-plasma QRT. The FOTC to the rate of second harmonic emission is derived in Sec.~IV, where its importance in various situations is also analyzed.

\section{Theoretical context}


Nonlinear plasma response tensors are defined by expanding an induced plasma property, such as the induced current density, in powers of the amplitude of the Fourier transformed electromagnetic field; this is termed the weak-turbulence expansion. The QRT is the coefficient of the second-order term in this expansion. On choosing to describe the electromagnetic field by the vector potential $\bm{A}$ in the temporal gauge, the induced current density is given in Fourier space by (e.g., Ref.~\onlinecite{melrose80b})
\begin{equation}
\label{eq:Ji}
J_i(k)=\sum_{n=1}^{\infty} J_i^{(n)}(k),
\end{equation}
where
\begin{align}
J_i^{(n)}(k)=&\int d\lambda^{(n)} \alpha_{i j_1 j_2 \cdots j_n}(k,k_1,\ldots,k_n) \nonumber \\
& \times A_{j_1}(k_1) A_{j_2}(k_2) \cdots A_{j_n}(k_n). \label{eq:Jin}
\end{align}
In Eqs.~(\ref{eq:Ji}) and (\ref{eq:Jin}), $k_m$ collectively denotes $\omega_m$ and $\bm{k}_m$ for the $m$th wave, and $d\lambda^{(n)}$ is the $n$th-order convolution integral given by 
\begin{align}
d\lambda^{(n)}=&\frac{d^4k_1}{(2\pi)^4}\frac{d^4k_2}{(2\pi)^4}\cdots\frac{d^4k_n}{(2\pi)^4} \nonumber \\
& \times (2\pi)^4\delta^4(k-k_1-\cdots-k_n), \label{eq:conv}
\end{align}
with
\begin{equation}
\label{eq:d4k}
d^4k=d\omega\,d^3\bm{k}
\end{equation}
and
\begin{equation}
\label{eq:delta4k}
\delta^4(k)=\delta(\omega)\delta^3(\bm{k}).
\end{equation}
The nonlinear response tensors are most commonly calculated via the Vlasov equation, which relates the distribution function to the wave fields for a collisionless plasma. Solving the Vlasov equation by employing a weak-turbulence expansion of the distribution function and expressing the induced current as a moment of the distribution yields the general form of the QRT (e.g., Ref.~\onlinecite{percival98a}),
\begin{fleqn}
\begin{align}
\label{eq:exact}
\alpha_{ijl}(k,k_1,k_2) = &\frac{q^3}{2m^2}\biggl\{ \left[k_i + \left(|\bm{k}|^2-\omega^2/c^2\right)\frac{\partial}{\partial k_i}\right] \nonumber \\
& \times [I(k)\delta_{jl}+k_{2j}J_l(k,k_2)+k_{1l}J_j(k,k_1) \nonumber \\
& +(\bm{k}_1\cdot\bm{k}_2 - \omega_1\omega_2/c^2)K_{jl}(k,k_1,k_2)] \nonumber \\
& + (i,k)\leftrightarrow (j,k_1) + (i,k)\leftrightarrow (l,k_2)\biggr\}.
\end{align}
where $(i,k)\leftrightarrow (j,k_1)$ represents the additional terms generated from those written explicitly by interchanging $k$ and $k_1$ and the associated tensor indices, and the integrals $I(k)$, $J_i(k_1,k_2)$, and $K_{ij}(k,k_1,k_2)$ are given by
\begin{align}
& I(k) = \int d^3\bm{v}\,f(\bm{v})\,\frac{1}{\omega-\bm{k}\cdot\bm{v}}, \label{eq:i} \\
& J_i(k_1,k_2) = \int d^3\bm{v}\,f(\bm{v})\,\frac{v_i}{(\omega_1-\bm{k}_1\cdot\bm{v})(\omega_2-\bm{k}_2\cdot\bm{v})}, \label{eq:j} \\
& K_{ij}(k,k_1,k_2) = \int d^3\bm{v}\,f(\bm{v}) \nonumber \\
& \times \frac{v_i v_j}{(\omega-\bm{k}\cdot\bm{v})(\omega_1-\bm{k}_1\cdot\bm{v})(\omega_2-\bm{k}_2\cdot\bm{v})}. \label{eq:k}
\end{align}
It was noted in Ref.~\onlinecite{percival98a} that the partial derivative $\partial K_{jl}/ \partial k_i$ cannot be taken without ambiguity since one of the wave vectors in the integrand is no longer independent of the other two. The integral
\begin{align}
\label{eq:l}
L_{ijl} &=\frac{\partial K_{jl}}{\partial k_i} \nonumber \\
&= \int d^3\bm{v}\,f(\bm{v}) \frac{v_i v_j v_l}{(\omega-\bm{k}\cdot\bm{v})^2(\omega_1-\bm{k}_1\cdot\bm{v})(\omega_2-\bm{k}_2\cdot\bm{v})}
\end{align}
\end{fleqn}
must therefore be evaluated directly.

Thermal effects in the wave coupling are often ignored for interactions between three fast waves; neglecting these effects is known as the cold-plasma approximation. The cold-plasma QRT may be calculated by substituting $f(\bm{v})=n\delta^3(\bm{v})$ into Eqs.~(\ref{eq:i})-(\ref{eq:l}) then evaluating the integrals. The integrals $J$, $K$, and $L$ vanish, and $I(k)$ is replaced by $n/\omega$. This leads to
\begin{equation}
\label{eq:cold}
\alpha_{ijl}^{(\mathrm{cold})}(k,k_1,k_2)=\frac{q^3n}{2m^2}\left(\frac{k_i\delta_{jl}}{\omega}+\frac{k_{1j}\delta_{il}}{\omega_1}+\frac{k_{2l}\delta_{ij}}{\omega_2}\right).
\end{equation}
The cold-plasma approximation is made in order to simplify the mathematical analysis, but the neglected thermal effects may become significant in some circumstances.

Percival has calculated the integrals in Eqs.~(\ref{eq:i})-(\ref{eq:l}) exactly for a plasma in which the particles have a Maxwellian velocity distribution.\cite{percival98a} This analysis yields an expression for the QRT in terms of generalized plasma dispersion functions,\cite{percival98b} but calculating the interaction rate (as described below) is problematic because it involves integrals of the squared modulus of the QRT contracted with the relevant polarization tensors. Due to the large number of terms in the exact QRT and the possible occurrence of catastrophic cancellation it is infeasible to use this response tensor directly in the calculation of rates.

Once the response tensors are known, the emission and absorption of waves can be studied. These processes are often treated semiclassically;\cite{tsytovich70,melrose80b} the waves are interpreted as a collection of wave quanta with momentum $\hbar\bm{k}$ and energy $\hbar|\omega_M(\bm{k})|$. The occupation number $N_M(\bm{k})$ is introduced, being defined as the number density of wave quanta within the elemental range $d^3\bm{k}$ of $\bm{k}$. This quantity is related to the energy density per unit volume of $\bm{k}$-space, $W_M(\bm{k})$, by
\begin{equation}
N_M(\bm{k})=\frac{W_M(\bm{k})}{\hbar\omega_M(\bm{k})}.
\end{equation}
In a coalescence process the current $\bm{J}^{(2)}(k)$ induced by the simultaneous response of the plasma to two wave fields $\bm{A}_P(k_1)$ and $\bm{A}_Q(k_2)$, given by Eq.~(\ref{eq:Jin}), is the source of a third wave field $\bm{A}_M(k)$, where $k=k_1+k_2$ as implied by the delta function. Assuming the random phase approximation, the rate equation for the wave mode $M$ in the three-wave interaction $P(k_1)+Q(k_2)\rightarrow M(k)$ is given by (e.g., Refs.~\onlinecite{tsytovich70} and \onlinecite{melrose80b})
\begin{align}
\frac{dN_M}{dt}(\bm{k}) = &\int \frac{d^3\bm{k}_1}{(2 \pi)^3} \int \frac{d^3\bm{k}_2}{(2 \pi)^3} u_{MPQ}(\bm{k},\bm{k}_1,\bm{k}_2) \nonumber \\
& \times \{N_P(\bm{k}_1)N_Q(\bm{k}_2) - N_M(\bm{k})\nonumber \\
& \times [N_P(\bm{k}_1)+N_Q(\bm{k}_2)]\}.
\end{align}
Alternatively, the rate can be expressed in terms of $T_M$, the effective temperature for the wave mode $M$, as
\begin{align}
\frac{dT_M}{dt}(\bm{k}) = & \int \frac{d^3\bm{k}_1}{(2 \pi)^3} \int \frac{d^3\bm{k}_2}{(2 \pi)^3} u_{MPQ}(\bm{k},\bm{k}_1,\bm{k}_2) \nonumber \\
& \times \{T_P(\bm{k}_1)T_Q(\bm{k}_2) - T_M(\bm{k})\nonumber \\
& \times [T_P(\bm{k}_1)+T_Q(\bm{k}_2)]\} \nonumber \\
& \times \omega_M(\bm{k})/(\hbar\omega_P(\bm{k}_1)\omega_Q(\bm{k}_2)),
\end{align}
where the effective temperature is related to the occupation number by
\begin{equation}
T_M(\bm{k})=\hbar\omega_M(\bm{k})N_M(\bm{k}).
\end{equation}
The equation for the interaction probability $u_{MPQ}$ is (e.g., Refs.~\onlinecite{tsytovich70} and \onlinecite{melrose80b})
\begin{align}
\label{eq:uMPQ}
u_{MPQ}(k,k_1,k_2)=&\frac{4\hbar}{\epsilon_0^3} \frac{R_M(\bm{k})R_P(\bm{k_1})R_Q(\bm{k_2})}{|\omega_M(\bm{k})\omega_P(\bm{k_1})\omega_Q(\bm{k_2})|}
\nonumber \\
&\times\vert\alpha_{MPQ}(k_M,k_{P1},k_{Q2})\vert^2 \nonumber \\
&\times\left(2\pi\right)^4 \delta^4(k_M-k_{P1}-k_{Q2}),
\end{align}
where $R_M$ is the ratio of electric to total energy in the wave mode $M$, and 
\begin{align}
\alpha_{MPQ}(k_M,k_{P1},k_{Q2}) = &\alpha_{ijl}(k_M,k_{P1},k_{Q2}) \nonumber \\
& \times e^{*}_{Mi}(\bm{k}) e_{Pj}(\bm{k}_1) e_{Ql}(\bm{k}_2),
\end{align}
with $\bm{e}_M(\bm{k})$ the polarization vector for the wave mode $M$. The quantity $k_M$ collectively denotes $\omega_M(\bm{k})$ and $\bm{k}$, and similarly for $k_{P1}$ and $k_{Q2}$. In Eq.~(\ref{eq:uMPQ}), the delta function in wave vector is interpreted in the semiclassical description as expressing conservation of momentum,
\begin{equation}
\bm{k}=\bm{k}_1+\bm{k}_2,
\end{equation}
and the delta function in frequency expresses conservation of energy,
\begin{equation}
\omega_M(\bm{k})=\omega_P(\bm{k_1})+\omega_Q(\bm{k_2}),
\end{equation}
where the common factor of $\hbar$ is omitted. 

\section{First-order thermal correction to the cold-plasma quadratic response tensor}

In this section, the general quadratic response tensor (QRT) in Eq.~(\ref{eq:exact}) is approximated by deriving the first-order thermal correction (FOTC) to the cold-plasma QRT given by Eq.~(\ref{eq:cold}). We discuss the expression obtained for the FOTC, including its importance and validity, for different interactions.

\subsection{Derivation}

First the resonant denominators in Eqs.~(\ref{eq:i})-(\ref{eq:l}) are binomially expanded in powers of $\bm{k}_m\cdot\bm{v}/\omega_m$ (which may be expressed as $v_{\parallel}/v_{\phi m}$ where $\parallel$ is with respect to $\bm{k}_m$) using the binomial expansion $(1-x)^{-1}=\sum_{n=0}^{\infty} x^n$. The $v_{\phi m}$ in the denominator of the expanded quantity ensures that this expansion of the integrals is convergent, as shown below. We show that this expansion recovers the cold-plasma QRT plus additional terms of order $(V/v_{\phi})^{2n}$ relative to the cold-plasma terms, where $v_{\phi}$ is the phase speed, $V$ is the thermal speed of the particles, and $n$ is an integer. The $n=1$ terms are called the FOTC to the cold-plasma QRT.

Two additional assumptions are made to derive the FOTC: the first is that the distribution function is isotropic, and the second is that the thermal effects for each of the three wave fields can be treated equally, which will be discussed after the derivation.

Firstly, we perform a binomial expansion of the resonant denominators of the integrals given by Eqs.~(\ref{eq:i})-(\ref{eq:l}). This gives
\begin{fleqn}
\begin{align}
\label{eq:ib}
I(k) = &\frac{1}{\omega}\int d^3\bm{v} f(\bm{v}) \Biggl[1+\frac{\bm{k}\cdot\bm{v}}{\omega}+\left(\frac{\bm{k}\cdot\bm{v}}{\omega}\right)^2 \nonumber \\
	&+O\left(\frac{\bm{k}\cdot\bm{v}}{\omega}\right)^3\Biggr], \\
\approx &\frac{1}{\omega}\int d^3\bm{v} f(\bm{v})+\frac{k_q}{\omega^2}\int d^3\bm{v} f(\bm{v})v_q \nonumber \\
	&+\frac{k_r k_s}{\omega^3}\int d^3\bm{v} f(\bm{v}) v_r v_s,
\end{align}
\begin{align}
J_i(k,k_1) = &\frac{1}{\omega \omega_1} \int d^3\bm{v} f(\bm{v}) v_i \Biggl[1+\frac{\bm{k}\cdot\bm{v}}{\omega}+\frac{\bm{k}_1\cdot\bm{v}}{\omega_1}
\nonumber \\
	&+O\left(\frac{\bm{k}\cdot\bm{v}}{\omega}\right)^2\Biggr], \\
\approx &\frac{1}{\omega \omega_1} \int d^3\bm{v} f(\bm{v}) v_i 
	+\frac{1}{\omega \omega_1} \left(\frac{k_s}{\omega}+\frac{k_{1s}}{\omega_1}\right) \nonumber \\
	&\times \int d^3\bm{v} f(\bm{v}) v_i v_s,
\label{eq:jb}
\end{align}
\begin{align}
K_{ij}(k,k_1,k_2) = &\frac{1}{\omega\omega_1\omega_2} \int d^3\bm{v} f(\bm{v}) v_i v_j \left[1+O\left(\frac{\bm{k}\cdot\bm{v}}{\omega}\right)\right] \\
\approx &\frac{1}{\omega\omega_1\omega_2} \int d^3\bm{v} f(\bm{v}) v_i v_j, \\
\label{eq:kb}
\end{align}
\begin{align}
L_{ijl}(k,k_1,k_2) = &\frac{1}{\omega\omega_1\omega_2} \int d^3\bm{v} f(\bm{v}) v_i v_j v_l \Biggl[1 \nonumber \\
& + O\left(\frac{\bm{k}\cdot\bm{v}}{\omega}\right)\Biggr] \\
\approx &\frac{1}{\omega\omega_1\omega_2} \int d^3\bm{v} f(\bm{v}) v_i v_j v_l.
\label{eq:lb}
\end{align}
\end{fleqn}

We let $v_i=\delta_{iq}v_q$ and $v_i=\delta_{ir}v_r$ in the first and second integrals in Eq.~(\ref{eq:jb}) respectively. Then substituting the binomially approximated $I$, $J$, $K$, and $L$ given by Eqs.~(\ref{eq:ib})-(\ref{eq:lb}) respectively into the general QRT in Eq.~(\ref{eq:exact}) gives
\begin{widetext}
\begin{fleqn}
\begin{align}
\alpha_{ijl}=&\frac{q^3}{2m^2}\Biggl\{ \Biggl[ k_i + \left(|\bm{k}|^2-\omega^2/c^2\right)\frac{\partial}{\partial k_i}\Biggr]
	\Biggl[\frac{\delta_{jl}}{\omega}\int d^3\bm{v} f(\bm{v})
 + \left( \frac{k_q \delta_{jl}}{\omega^2}+\frac{k_{2j}\delta_{lq}}{\omega\omega_2} + \frac{k_{1l}\delta_{jq}}{\omega\omega_1}\right) 
	A_q(\bm{v}) \nonumber \\
& + \biggl\{ \frac{k_r k_s \delta_{jl}}{\omega^3} + \frac{k_{2j}\delta_{lr}}{\omega\omega_2}\left(\frac{k_s}{\omega}+\frac{k_{2s}}{\omega_2}\right)
	+ \frac{k_{1l}\delta_{jr}}{\omega\omega_1}\left(\frac{k_s}{\omega}+\frac{k_{1s}}{\omega_1}\right)\biggr\}B_{rs}(\bm{v})\Biggr] 
	+ \frac{\left(\bm{k}_1\cdot\bm{k}_2-\omega_1\omega_2/c^2\right)}{{\omega\omega_1\omega_2}} \nonumber \\
& \times \Bigl[k_i B_{jl}(\bm{v}) + \left(|\bm{k}|^2-\omega^2/c^2\right) C_{ijl}(\bm{v})\Bigr]
 + (i,k)\leftrightarrow (j,k_1) + (i,k)\leftrightarrow (l,k_2)\Biggr\}.
\end{align}
\end{fleqn}
\end{widetext}
Here we have defined
\begin{align}
&A_i(\bm{v})=\int d^3\bm{v}f(\bm{v})v_i, \\
&B_{ij}(\bm{v})=\int d^3\bm{v}f(\bm{v})v_i v_j, \\
&C_{ijl}(\bm{v})=\int d^3\bm{v}f(\bm{v})v_i v_j v_l.
\end{align}

We now reexpress the integrals over velocity space in terms of moments of the distribution function $f(\bm{v})$. The distribution function is defined such that the particle number density $n$ is given by 
\begin{equation}
\label{eq:n}
n=\int d^3\bm{v} f(\bm{v}).
\end{equation}
The moment of a quantity $Q(\bm{v})$ is defined by
\begin{equation}
\label{eq:moment}
\langle Q(\bm{v}) \rangle = \frac{1}{n}\int d^3\bm{v} f(\bm{v}) Q(\bm{v}).
\end{equation}
We make the assumption that the distribution function is isotropic, hence $f(\bm{v})=f(v)$  where $v=(v_x^2+v_y^2+v_z^2)^{1/2}$ and so $f$ is an even function of $v_x$, $v_y$, and $v_z$. In Cartesian coordinates, the tensor indices run over $x$, $y$, and $z$. The integrals $A_i$ and $C_{ijl}$ then vanish because every choice of indices gives an integrand that is odd in one or all of the variables $v_x$, $v_y$, and $v_z$, and the integration limits are symmetric about the origin. By symmetry in $v_x$, $v_y$, and $v_z$ all the diagonal components of $B_{ij}$ are equal, as are the off-diagonal components. The off-diagonal components of the $B_{ij}$ vanish due to oddness of the integrand, so $B_{ij} = B\delta_{ij}$ where $B=B_{xx}=B_{yy}=B_{zz}$. To calculate $B$ we note that $B=(B_{xx}+B_{yy}+B_{zz})/3$ and so
\begin{align}
B_{ij} = &\frac{\delta_{ij}}{3}\int d^3\bm{v} f(v) (v_x^2+v_y^2+v_z^2), \\
= &\frac{n\langle v^2 \rangle \delta_{ij}}{3} \label{eq:B}
\end{align}
from Eq.~(\ref{eq:moment}). 

Substituting Eqs.~(\ref{eq:n}) and (\ref{eq:B}), and $A_q=C_{ijl}=0$, into Eq.~(\ref{eq:exact}) and simplifying gives
\begin{fleqn}
\begin{align}
\label{eq:natunits}
\alpha_{ijl}=&\frac{q^3n}{2m^2}\Biggl\{\Biggl[ k_i + \left(|\bm{k}|^2-\omega^2/c^2\right)\frac{\partial}{\partial k_i}\Biggr]
	\Biggl[\frac{\delta_{jl}}{\omega} + \frac{\langle v^2 \rangle}{3} \nonumber \\
&  \times \biggl\{ \frac{|\bm{k}|^2 \delta_{jl}}{\omega^3} + \frac{k_{2j}}{\omega\omega_2}\left(\frac{k_l}{\omega}+
	\frac{k_{2l}}{\omega_2}\right) + \frac{k_{1l}}{\omega\omega_1} \nonumber \\
& \times \left(\frac{k_j}{\omega}+\frac{k_{1j}}{\omega_1}\right)\biggr\}\Biggr] + \frac{\langle v^2 \rangle \delta_{jl}}{3} \frac{k_i\left(\bm{k}_1\cdot\bm{k}_2-\omega_1\omega_2/c^2\right)}{\omega\omega_1\omega_2} \nonumber \\
& + (i,k)\leftrightarrow (j,k_1) + (i,k)\leftrightarrow (l,k_2)\Biggr\},
\end{align}
\end{fleqn}
where the arguments of the response tensor have been omitted for brevity. On performing the derivative in Eq.~(\ref{eq:natunits}), using $\partial|\bm{k}|^2/\partial k_i = 2k_i$ and $\partial k_j/\partial k_i=\delta_{ij}$, we have
\begin{fleqn}
\begin{align}
\label{eq:stuff}
\alpha_{ijl}=&\frac{q^3n}{2m^2}\Biggl\{ \frac{k_i\delta_{jl}}{\omega} 
	+ \frac{\langle v^2 \rangle}{3} \biggl[ k_i \biggl\{ \frac{|\bm{k}|^2 \delta_{jl}}{\omega^3} + \frac{k_{2j}}{\omega\omega_2}\left(\frac{k_l}{\omega}
	+ \frac{k_{2l}}{\omega_2}\right) \nonumber \\
& + \frac{k_{1l}}{\omega\omega_1}\left(\frac{k_j}{\omega}+\frac{k_{1j}}{\omega_1}\right)\biggr\}\frac{\left(\bm{k}_1\cdot\bm{k}_2-\omega_1\omega_2/c^2\right)k_i\delta_{jl}}{\omega\omega_1\omega_2}
	\nonumber \\
& + \left(|\bm{k}|^2-\omega^2/c^2\right) \left(\frac{2k_i\delta_{jl}}{\omega^3}+\frac{k_{2j}\delta_{il}}{\omega^2\omega_2}
	+ \frac{k_{1l}\delta_{ij}}{\omega^2\omega_1}\right) \biggr] \nonumber \\
& + (i,k)\leftrightarrow (j,k_1) + (i,k)\leftrightarrow (l,k_2)\biggr\}.
\end{align}
\end{fleqn}
Rearrangement and factorization of Eq.~(\ref{eq:stuff}) yields
\begin{fleqn}
\begin{align}
\label{eq:rel}
\alpha_{ijl}=&\frac{q^3n}{2m^2}\biggl\{ \frac{k_i\delta_{jl}}{\omega} 
	+ \frac{\langle v^2 \rangle}{3} \biggl[\biggl( \frac{|\bm{k}|^2+2(|\bm{k}|^2-\omega^2/c^2)}{\omega^2} \nonumber \\
& + \frac{\bm{k}_1\cdot\bm{k}_2-\omega_1\omega_2/c^2}{\omega_1\omega_2}\biggr)\frac{k_i\delta_{jl}}{\omega} + \frac{|\bm{k}|^2-\omega^2/c^2}{\omega^2}
	\biggr(\frac{k_{1l}\delta_{ij}}{\omega_1} \nonumber \\
& +\frac{k_{2j}\delta_{il}}{\omega_2}\biggr)
	+ \frac{k_i}{\omega} \left( \frac{k_{1l} k_j}{\omega \omega_1} + \frac{k_{1l} k_{1j}}{\omega_1^2} + \frac{k_{2j} k_l}{\omega \omega_2} 
	+ \frac{k_{2j} k_{2l}}{\omega_2^2}\right)\biggr] \nonumber \\
& + (i,k)\leftrightarrow (j,k_1) + (i,k)\leftrightarrow (l,k_2)\biggr\}.
\end{align}
\end{fleqn}
The first term inside the braces and its interchanges are identified as the cold-plasma QRT given by Eq.~(\ref{eq:cold}); the remaining terms are the 
FOTC to Eq.~(\ref{eq:cold}), denoted by $\Delta\alpha_{ijl}$. That is, $\alpha_{ijl}=\alpha_{ijl}^{(\mathrm{cold})}+\Delta\alpha_{ijl}$. In the nonrelativistic ($c\rightarrow \infty$) limit, the FOTC to the cold-plasma QRT, for an isotropic particle velocity distribution, is
\begin{align}
\label{eq:FOTC}
\Delta\alpha_{ijl}=&\frac{q^3n}{2m^2}\frac{\langle v^2 \rangle}{3}\biggl\{ \biggl[ \biggl( \frac{3|\bm{k}|^2}{\omega^2}
	+ \frac{\bm{k}_1\cdot\bm{k}_2}{\omega_1\omega_2}\biggr)\frac{k_i\delta_{jl}}{\omega} + \frac{|\bm{k}|^2}{\omega^2} \nonumber \\
& \times \biggl(\frac{k_{1l}\delta_{ij}}{\omega_1}+\frac{k_{2j}\delta_{il}}{\omega_2}\biggr)
	+ \frac{k_i}{\omega} \biggl( \frac{k_{1l} k_j}{\omega \omega_1} + \frac{k_{1l} k_{1j}}{\omega_1^2} \nonumber \\
& + \frac{k_{2j} k_l}{\omega \omega_2} 	+ \frac{k_{2j} k_{2l}}{\omega_2^2}\biggr)\biggr] \nonumber \\
& + (i,k)\leftrightarrow (j,k_1) + (i,k)\leftrightarrow (l,k_2)\biggr\}.
\end{align}

\subsection{Discussion}

The particle temperature $T$ is related to the second moment of the distribution function by the definition $k_B T=m\langle v^2\rangle /3$, where $k_B$ is Boltzmann's constant. For a Maxwellian distribution
\begin{equation}
f(\bm{v})=\frac{n}{(2\pi)^{3/2}V^3}e^{-v^2/2V^2},
\end{equation}
the thermal speed is given by $V=\sqrt{k_B T/m}$, whence $\langle v^2\rangle /3$ is replaced by $V^2$ in Eq.~(\ref{eq:FOTC}); this reproduces the expression for $\Delta\alpha_{ijl}$ in Ref.~\onlinecite{percival_thesis}.

The terms in the cold-plasma QRT and FOTC are of order $1/v_{\phi}$ and $V^2/v_{\phi}^3$ respectively, from Eqs.~(\ref{eq:cold}) and (\ref{eq:FOTC}),
where $v_{\phi}$ here represents the phase speeds of any of the three interacting waves. The ratio of these orders, $V^2/v_{\phi}^2$, therefore determines the significance of the FOTC: if it is much less than unity then the cold plasma approximation will be accurate, but as it approaches unity  the cold plasma approximation begins to break down and the FOTC must be included for an accurate description of the plasma response.

From Eqs.~(\ref{eq:exact}) and (\ref{eq:ib})-(\ref{eq:lb}), it follows that higher-order thermal corrections to the cold-plasma QRT involve integrals of the form
\begin{equation}
S_{i_1 i_2\cdots i_n}(\bm{v})=\int d^3\bm{v} f(v) v_{i_1} v_{i_2} \cdots v_{i_n},
\end{equation}
where $i_m$ are tensor indices running over $x$, $y$, and $z$. For odd $n$, each choice of indices will give an odd power in at least one of the variables $v_x$, $v_y$, or $v_z$, hence in this case the integral will be zero. Therefore, the $n$th-order thermal correction will be 
$O\left\{(V/v_{\phi})^{2n}\right\}$ relative to the cold-plasma terms. For the expansion of the general QRT by a binomial expansion of the resonant denominators to be convergent, one requires $v_{\phi}>V$ for each wave mode.

Thermal effects from one or two of the participating waves may be more important than those from the other wave or waves. This is the situation for the Langmuir-wave coalescence process: although both Langmuir and transverse waves are fast, the thermal effects from the Langmuir waves will be more important since they have a significantly lower phase speed than the transverse wave, and so the terms involving the transverse wave phase speed may be neglected. Equation (\ref{eq:FOTC}) may also be applied to the Raman scattering process $L+T\rightarrow T'$ which has been proposed for third and higher harmonic emission.\cite{zlotnik78,cairns87b,rhee09} However, the FOTC will not be as important as in Langmuir-wave coalescence since Raman scattering involves two transverse waves and only one Langmuir wave.
 
\section{First-order thermal correction to the second harmonic emission rate} 

In this section the FOTC to the cold-plasma QRT, given by Eq.~(\ref{eq:FOTC}), is applied to the rate calculation of second harmonic plasma emission via Langmuir-wave coalescence. The ratio of the FOTC rate to the cold-plasma rate is derived, and its range of values is calculated for different physical situations to assess the importance of the FOTC.

\subsection{Derivation}

The FOTC to quantities in this section will be denoted by the prefix $\Delta$, such that $x\approx x^{(\mathrm{cold})}+\Delta x$ for some quantity $x$. We define the electron thermal speed by $v_e=\sqrt{\langle v^2 \rangle/3}$, where the angle brackets denote the moment of the electron distribution function. Primary Langmuir waves $L(k_1)$ are assumed to be generated by an electron beam via a bump-on-tail instability, such that the phase speed of the Langmuir waves is approximately equal to the speed of the electron beam, i.e.~$v_{\phi 1}\approx v_b$. Backscattered Langmuir waves $L'(k_2)$, with which the primary waves coalesce, are assumed to be the product of the decay process $L\rightarrow L'+S$.\cite{melrose82,cairns87a,robinson98a,robinson98b,robinson98c,li05} Since the mass of the ions is much greater than that of the electrons and the QRT has a $m^{-2}$ dependence for each particle species, the ionic contribution to the QRT is neglected.

We first outline the derivation of the cold-plasma interaction probability for Langmuir-wave coalescence (see e.g. Ref~\onlinecite{melrose80b}). Transverse waves have $\bm{e}_T\cdot\bm{\kappa}_T=0$ and hence $\bm{e}^*_T\cdot\bm{\kappa}^*_T=0$. In the case of no spatial damping, $\bm{\kappa}_T$ is real and so $\bm{e}^*_T\cdot\bm{\kappa}_T=0$. On contracting the cold-plasma QRT in Eq.~(\ref{eq:cold}) with 
$e^*_{Ti}(\bm{k_T}) e_{Lj}(\bm{k_1}) e_{L'l}(\bm{k_{2}})$, i.e.~with $e^*_{Ti} \kappa_{1j} \kappa_{2l}$, where $\bm{\kappa}=\bm{k}/|\bm{k}|$, we have
\begin{equation}
\alpha_{TLL'}^{(\mathrm{cold})}(k_T,k_1,k_2) = \frac{e^3 n_e}{2 m_e^2}\left( \frac{\edotk{1}}{v_{\phi 2}} + \frac{\bm{e}^*_T \cdot
\bm{\kappa}_{2}}{v_{\phi 1}} \right);
\end{equation}
hence, using $|x+y|^2=|x|^2+|y|^2+2\mathrm{Re}\{x^*y\}$, we find
\begin{align}
\left|\alpha_{TLL'}^{(\mathrm{cold})}\right|^2 = &\frac{e^6 n_e^2}{4m_e^4}\biggl(\frac{|\edotk{1}|^2}{v_{\phi 2}^2}
+\frac{|\edotk{2}|^2}{v_{\phi 1}^2} \nonumber \\
& + 2\mathrm{Re}\left[\frac{(\edotk{1})^*(\edotk{2})}{v_{\phi 1}v_{\phi 2}}\right]\biggr).
\end{align}
When the polarization of the transverse waves $T$ is of no interest, an average over the two initial states of polarization and a sum over the two final states of polarization is performed. This leads to the replacement
\begin{equation}
\left|\edotk{1}\right|^2=\frac{1}{2}\left|\bm{\kappa}_T\times\bm{\kappa}_1\right|^2
\end{equation}
and similarly for $|\edotk{2}|^2$. Making the approximations that 
$R_L \approx \frac{1}{2},\, \omega_L \approx \omega_p,\,\mathrm{and}\,\,\omega_T \approx 2\omega_p$, and since $R_T=\frac{1}{2}$, we have the interaction probability for a cold plasma as
\begin{equation}
\label{eq:uMPQcold}
u_{TLL'}^{(\mathrm{cold})} \approx \frac{\hbar}{4\epsilon_0^3\omega_p^3} \left|\alpha_{TLL'}^{(\mathrm{cold})}\right|^2 (2\pi)^4 \delta^4(k_T-k_1-k_2).
\end{equation}

To obtain $\Delta\alpha_{TLL'}$, Eq.~(\ref{eq:FOTC}) is contracted with $e^*_{Ti} \kappa_{1j} \kappa_{2l}$. Grouping $\Delta\alpha_{TLL'}$ by order in $v_{\phi T}$ gives
\begin{equation}
\Delta\alpha_{TLL'} = \Delta\alpha_{TLL'}^{(0)} + \Delta\alpha_{TLL'}^{(1)} + \Delta\alpha_{TLL'}^{(2)}
\end{equation}
where
\begin{fleqn}
\begin{align}
\Delta\alpha_{TLL'}^{(0)} =& \frac{e^3 n_e}{2 m_e^2} v_e^2 \biggl[ \frac{3\,\edotk{2}}{v_{\phi 1}^3} + \frac{2(\edotk{2})(\bm{\kappa}_1 \cdot \bm{\kappa}_{2})}{v_{\phi 1}^2 v_{\phi 2}} \nonumber \\
	& + \frac{2(\edotk{1})(\bm{\kappa}_1 \cdot \bm{\kappa}_{2})}{v_{\phi 1} v_{\phi 2}^2} + \frac{3\,\bm{e}^*_T \cdot
	\bm{\kappa}_{1}}{v_{\phi 2}^3} \biggr], \\
\Delta\alpha_{TLL'}^{(1)} =& \frac{e^3 n_e}{2 m_e^2} \frac{v_e^2}{v_{\phi T}} \biggl[ 
	\frac{2(\bm{\kappa}_T\cdot\bm{\kappa}_{2})(\edotk{1})}
	{v_{\phi1}^2} \nonumber \\
	& + \frac{(\bm{\kappa}_T\cdot\bm{\kappa}_{2})(\edotk{2})+(\bm{\kappa}_T\cdot\bm{\kappa}_{1})
	(\edotk{1})}{v_{\phi 1}v_{\phi 2}} \nonumber \\ 
	& + \frac{2(\bm{\kappa}_T\cdot\bm{\kappa}_{1})(\edotk{2})}{v_{\phi 2}^2} \biggr], \\
\Delta\alpha_{TLL'}^{(2)} =& \frac{e^3 n_e}{2 m_e^2} \frac{v_e^2}{v_{\phi T}^2} \biggl[
	\frac{(\bm{\kappa}_1\cdot\bm{\kappa}_{2})(\edotk{1})}{v_{\phi 1}} \nonumber \\
	& + \frac{(\bm{\kappa}_1\cdot\bm{\kappa}_{2})(\edotk{2})}{v_{\phi 2}} \biggr].
\end{align}
\end{fleqn}

From these equations, $\Delta\alpha_{TLL'}^{(0)} = O\left(v_e^2/v_b^3\right)$, $\Delta\alpha_{TLL'}^{(1)} = O\left(v_e^2/v_b^2 v_{\phi T}\right)$, and $\Delta\alpha_{TLL'}^{(2)} = O\left(v_e^2/v_b v_{\phi T}^2\right)$. The electron beam speed is typically less than a few tenths of the speed of light and $v_{\phi T}>c$, so to first order $\Delta\alpha_{TLL'} \approx \Delta\alpha_{TLL'}^{(0)}$.

Including the first order thermal correction in the nonlinear response implies $\alpha_{TLL'} = \alpha_{TLL'}^{(\mathrm{cold})}+\Delta\alpha_{TLL'}$. Hence, to the next order after the cold-plasma term,
\begin{equation}
\label{eq:a2}
|\alpha_{TLL'}|^2 = \left|\alpha_{TLL'}^{(\mathrm{cold})}\right|^2 + 2\,\mathrm{Re}\left[\left(\alpha_{TLL'}^{(\mathrm{cold})}\right)^*\Delta\alpha_{TLL'}^{(0)}\right].
\end{equation}
So, the second term on the right hand side of Eq.~(\ref{eq:a2}) is the first order correction to $|\alpha_{TLL'}|^2$, which is then
\begin{widetext}
\begin{fleqn}
\begin{align}
\label{eq:da2}
\Delta\left(|\alpha_{TLL'}|^2\right) = & \frac{e^6n_e^2v_e^2}{2m_e^2}\,\mathrm{Re} \biggl\{ \frac{3|\edotk{2}|^2}{v_{\phi 1}^4} + \frac{[2(\edotk{2})(\bm{\kappa}_1\cdot\bm{\kappa}_{2}) + \edotk{1}]\left(\edotk{2}\right)^*+3(\edotk{2})\left(\edotk{1}\right)^*}{v_{\phi 1}^3 v_{\phi 2}} \nonumber \\
& + \frac{[2(\edotk{2})(\bm{\kappa}_1\cdot\bm{\kappa}_{2}) + \edotk{1}]\left(\edotk{1}\right)^*}{v_{\phi 1}^2 v_{\phi 2}^2}
+(\bm{\kappa}_1,v_{\phi 1})\rightarrow(\bm{\kappa}_2,v_{\phi 2}) \biggr\}.
\end{align}
\end{fleqn}
\end{widetext}
From Eqs.~(\ref{eq:uMPQ}), (\ref{eq:a2}), and (\ref{eq:da2}), the FOTC to the interaction probability is
\begin{equation}
\label{eq:uMPQcold}
\Delta u_{TLL'} \approx \frac{\hbar}{4\epsilon_0^3\omega_p^3} \Delta\left(\left|\alpha_{TLL'}\right|^2\right) (2\pi)^4 \delta^4(k_T-k_1-k_2).
\end{equation}

The FOTC to the interaction rate is next calculated in terms of the effective temperature using a modified system of spherical coordinates (as in Ref.~\onlinecite{li05}), where the Langmuir wave numbers $k_{1,2}$ take on positive and negative values, and the polar angle $\theta$ ranges from $0$ to $\pi/2$. The angle between the primary and backscattered Langmuir waves is assumed to be greater than $\pi/2$, hence $\mathrm{sgn}[k_2]=-\mathrm{sgn}[k_1]$.
If the back-reaction $T\rightarrow L+L'$ is ignored, the FOTC to the rate of Langmuir-wave coalescence is given approximately by
\begin{align}
\Delta\frac{dT_T}{dt}(\bm{k}_T) = & \frac{2}{\hbar\omega_p} \int \frac{d^3\bm{k}_1}{(2 \pi)^3} \int \frac{d^3\bm{k}_2}{(2 \pi)^3} \nonumber \\
& \times \Delta u_{TLL'}(\bm{k}_T,\bm{k}_1,\bm{k}_2) \nonumber \\
& \times T_L(\bm{k}_1)T_{L'}(\bm{k}_2).
\end{align}
We assume that $\omega_{1,2}\approx\omega_p$, whence $v_{\phi 1,2}=\omega_p/k_{1,2}$, to simplify the integrand. The delta function $\delta^3(\bm{k}_T - \bm{k}_1 -\bm{k}_{2})$ is used to integrate over $d^3\bm{k}_{2}$. Thus on integration,
\begin{align}
&\bm{k}_{2} \rightarrow \bm{k}_T - \bm{k}_1, \\
&\edotk{2} \rightarrow \frac{-k_1}{k_2(\bm{k}_1,\bm{k}_T)}\edotk{1}, \\
&\bm{\kappa}_1\cdot\bm{\kappa}_{2} \rightarrow \frac{k_T\cos\psi-k_1}{k_2(\bm{k}_1,\bm{k}_T)},
\end{align}
where $\psi$ is the angle between the $T$ and $L$ wave vectors, and
\begin{equation}
k_2(\bm{k}_1,\bm{k}_T)=-\mathrm{sgn}[k_1]\left(k_1^2+k_T^2-2k_1k_T\cos\psi\right)^{1/2}.
\end{equation}
The rate then becomes
\begin{fleqn}
\begin{align}
\label{eq:T_rate}
\Delta\frac{dT_T}{dt}(\bm{k}_T) = &\frac{e^2 v_e^2}{16\pi^2\epsilon_0 m_e^2 \omega_p^8} 
\int d^3\bm{k}_1 g(\bm{k}_1,\bm{k}_T) \nonumber \\
& \times \delta\left[\omega_T(\bm{k}_T)-\omega_L(\bm{k}_1)-\omega_{L}(\bm{k}_T-\bm{k}_1)\right] \nonumber \\
& \times T_L(\bm{k}_1)T_{L'}(\bm{k}_T-\bm{k}_1)\left|\edotk{1}\right|^2,
\end{align}
\end{fleqn}
with
\begin{fleqn}
\begin{align}
\label{eq:f}
g(\bm{k}_1,\bm{k}_T)=&\frac{3k_1^6}{[k_2(\bm{k}_1,\bm{k}_T)]^2}+2k_1^4\left[\frac{k_1\left(k_T\cos\psi-k_1\right)}{[k_2(\bm{k}_1,\bm{k}_T)]^2} -2 \right] \nonumber \\ 
& + k_1^2 [k_2(\bm{k}_1,\bm{k}_T)]^2 \biggl[ 1-\frac{4k_1\left(k_T\cos\psi-k_1\right)}{[k_2(\bm{k}_1,\bm{k}_T)]^2} \nonumber \\
& + \frac{k_1^2}{[k_2(\bm{k}_1,\bm{k}_T)]^2} \biggr] + 2k_1[k_2(\bm{k}_1,\bm{k}_T)]^3 \nonumber \\
& \times \left[\frac{(k_T\cos\psi-k_1)-2k_1}{k_2(\bm{k}_1,\bm{k}_T)}\right] + 3[k_2(\bm{k}_1,\bm{k}_T)]^4.
\end{align}
\end{fleqn}
Expanding Eq.~(\ref{eq:f}), simplifying, and factorizing leads to
\begin{align}
g(\bm{k}_1,\bm{k}_T)=&\frac{\left\{k_1^2-[k_2(\bm{k}_1,\bm{k}_T)]^2\right\}^2}{{[k_2(\bm{k}_1,\bm{k}_T)]^2}} \nonumber \\
& \times \bigl\{3k_1^2+3[k_2(\bm{k}_1,\bm{k}_T)]^2+2k_1 \nonumber \\
& \times(k_T\cos\psi-k_1)\bigr\} \\
=& \frac{k_T^2 (2k_1\cos\psi - k_T)^2}{{[k_2(\bm{k}_1,\bm{k}_T)]^2}}  \nonumber \\
& \times (4k_1^2+3k_T^2-4k_1k_T\cos\psi). \label{eq:g}
\end{align}
On summing the states of polarization of the $T$ waves in the final state and averaging over the initial states of polarization, 
$|\edotk{1}|^2$ is replaced by $|\bm{\kappa}_T\times\bm{\kappa}_1|^2/2=(\sin^2\psi)/2$ in Eq.~(\ref{eq:T_rate}).

To simplify the delta function, the assumption
\begin{equation}
\omega_{L}(\bm{k}) \approx \omega_p + 3k^2v_e^2/2\omega_p
\label{eq:omegaL}
\end{equation}
is made for both Langmuir waves, which is valid for $k\ll \lambda_D^{-1}=(v_e/\omega_p)^{-1}$. The dispersion relation for Langmuir waves in a generalized-Lorentzian plasma (i.e., one in which the electrons have a kappa distribution) is also given by Eq.~(\ref{eq:omegaL}) in the limit $k\ll \lambda_D^{-1}$. \cite{thorne91} So, on substituting Eq.~(\ref{eq:omegaL}) into the delta function in Eq.~(\ref{eq:T_rate}), we find
\begin{align}
&\delta[\omega_T(\bm{k}_T) -\omega_L(\bm{k}_1)-\omega_L(\bm{k}_T-\bm{k}_1)] \nonumber \\
&=\delta\left\{\left[\omega_T(\bm{k}_T)-2\omega_p\right]-\frac{3v_e^2}{2\omega_p}\left(2k_1^2+k_T^2-2k_1k_T\cos\psi\right)\right\}.
\end{align}
Using $\delta(ax)=\delta(x)/|a|$ gives
\begin{align}
&\delta[\omega_T(\bm{k}_T) -\omega_L(\bm{k}_1)-\omega_L(\bm{k}_T-\bm{k}_1)] \nonumber \\
&=\frac{\omega_p}{3v_e^2|k_1|k_T}\delta\biggl\{\cos\psi-\frac{1}{2k_1k_T}\biggl[2k_1^2+k_T^2 \nonumber \\
& -\frac{2\omega_p}{3v_e^2}[\omega_T(\bm{k}_T)-2\omega_p]\biggr]\biggr\}.
\end{align}

Evaluating the integrals over $\cos\theta$ and $\phi$ leads to the rate equation
\begin{widetext}
\begin{align}
\Delta\frac{\partial T_T}{\partial t}(k_T,\chi) = & \frac{e^2}{48\pi\epsilon_0m_e^2\omega_p^3} \int dk_1\,g(k_1,k_T) \nonumber \\
& \times \exp\left\{\beta\cos\chi\left[\cos\psi(k_1,k_T)\left(1+\frac{|k_1|}{|k_2(k_1,k_T)|}\right)+\frac{k_T}{k_2(k_1,k_T)}\right]\right\} \nonumber \\
& \times I_0\left[\beta\sin\chi\left[1-\cos^2\psi(k_1,k_T)\right]^{1/2}\left(1+\frac{|k_1|}{|k_2(k_1,k_T)|}\right)\right]T_L(k_1)T_L[k_2(k_1,k_T)].
\label{eq:dTdt}
\end{align}
\end{widetext}
where $I_0$ is a modified Bessel function, $\chi$ is the angle between $\bm{k}_T$ and the $k_\parallel$ axis, and $\cos\psi$ satisfies
\begin{equation}
\label{eq:cospsi}
\cos\psi(k_1,k_T) = \frac{1}{2k_1k_T}\left(2k_1^2+k_T^2-\frac{2\omega_p}{3v_e^2}[\omega_T(k_T)-2\omega_p]\right).
\end{equation}
Substituting Eq.~(\ref{eq:cospsi}) into Eq.~(\ref{eq:g}) yields
\begin{fleqn}
\begin{align}
g(k_1,k_T)=&\frac{|k_1|k_T[2k_1\cos\psi(k_1,k_T)-k_T]^2}{[k_2(k_1,k_T)]^2} \nonumber \\
& \times \left(k_T^2+\frac{4\omega_p}{3v_e^2}\left[\omega_T(k_T)-2\omega_p\right]\right) \nonumber \\
& \times [1-\cos^2\psi(k_1,k_T)].
\end{align}
\end{fleqn}

The cold-plasma interaction rate derived by \citet{li05} is given by Eq.~(\ref{eq:dTdt}) on replacing $g(k_1,k_T)$ with
\begin{align}
h(k_1,k_T)=&\frac{\omega_p^2}{2v_e^2}\frac{|k_1|k_T[2k_1\cos\psi(k_1,k_T)-k_T]^2}{[k_2(k_1,k_T)]^2} \nonumber \\
& \times [1-\cos^2\psi(k_1,k_T)].
\end{align}
The ratio $R=g(k_1,k_T)/h(k_1,k_T)$ is then given by
\begin{equation}
\label{eq:ratio}
R(k_T)=\frac{2k_T^2v_e^2}{\omega_p^2} + \frac{8\left[\omega_T(k_T)-2\omega_p\right]}{3\omega_p}.
\end{equation}
Since $R$ is not a function of $k_1$, Eq.~(\ref{eq:ratio}) gives the ratio of the FOTC to the cold-plasma interaction rate, independent of the integral over $k_1$. On subsituting the minimum transverse wave number $k_{T0}=\omega_p\sqrt{3}/c$ into Eq.~(\ref{eq:R}), the ratio can be expressed as
\begin{equation}
R(k_T)=6\left(\frac{v_e}{c}\right)^2 \left(\frac{k_T}{k_{T0}}\right)^2 + \frac{8}{3}\left\{\left[1+3\left(\frac{k_T}{k_{T0}}\right)^2\right]^{\frac{1}{2}}-2\right\}.
\label{eq:R}
\end{equation}

To obtain the emission rate, a particular Langmuir spectrum $T_L(k)$ must first be assumed, after which the integral in Eq.~(\ref{eq:dTdt}) can be evaluated. For a Gaussian Langmuir spectrum,
\begin{equation}
T_L(k)=\exp\left[-\frac{(k-k_f)^2}{K_f^2}\right]+\exp\Biggl[-\frac{(k-k_b)^2}{K_b^2}\Biggr],
\label{eq:Teff}
\end{equation}
the emission rate peaks at a transverse wave number
\begin{equation}
\label{eq:ktmax}
k_{T\mathrm{max}}=k_{T0}(1+\epsilon)
\end{equation}
where $\epsilon = (k_f^2+k_b^2)/k_D^2$ with $k_D=\lambda_D^{-1}$. Since the $L'$ waves are produced by the electrostatic decay process $L\rightarrow L'+S$, $k_b \approx k_f-k_0$,\cite{cairns87a,willes96} where $k_f=\omega_p/v_b$ and $k_0=2\omega_pv_s/3v_e^2$ with $v_s$ the ion acoustic speed. This leads to
\begin{equation}
\label{eq:Delta}
\epsilon \approx 2\left(\frac{v_e}{v_b}\right)^2\left(1-\frac{2v_sv_b}{3v_e^2}+\frac{2v_s^2v_b^2}{9v_e^4}\right)
\end{equation}
in Eq.~(\ref{eq:ktmax}).

\subsection{Discussion}

The FOTC to the cold-plasma rate of second harmonic emission has been derived by applying the FOTC to the cold-plasma QRT. This derivation is valid for both Maxwellian and generalized Lorentzian distributions of electrons; this is due to the FOTC to the cold-plasma QRT being valid for arbitrary isotropic
velocity distributions, and to Langmuir waves having the same dispersion relation for both distributions in the long wavelength ($k\lambda_D\ll 1$) limit. The resulting ratio $R(k_T)$ of the FOTC to the cold-plasma emission rate, given by Eq.~(\ref{eq:R}), does not depend on the integration over $k_1$, so it is the same for all Langmuir wave spectra.

\begin{figure}
  \begin{center}
    \includegraphics{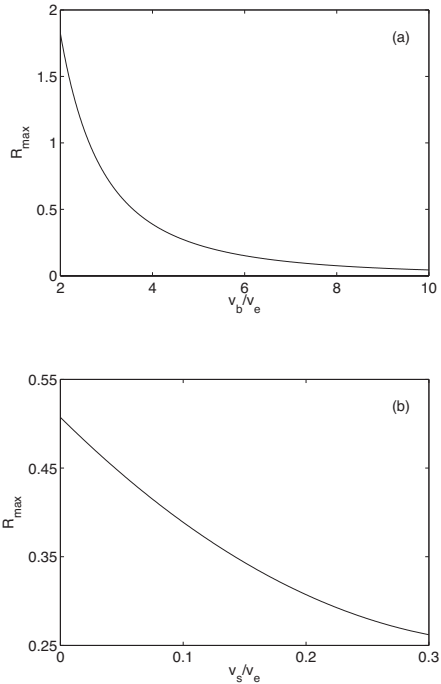}
  \end{center}
  \caption{$R_{\mathrm{max}}$ versus (a) $v_b/v_e$, where $v_s/v_e=0.1$; and (b) $v_s/v_e$, where $v_b/v_e=4$. For both (a) and (b), $v_e = 0.003c$.} 
  \label{fig:rmax}
\end{figure}

We define the quantity $R_{\mathrm{max}}$ to be the ratio $R(k_T)$ evaluated at $k_T=k_{T\mathrm{max}}$ in Eq.~(\ref{eq:R}). We choose $R_{\mathrm{max}}$ to quantify the importance of the FOTC to the second harmonic emission rate. For most applications $v_e/c\ll 1$, and so the first term in Eq.~(\ref{eq:R}) is small. Importantly, Eqs.~(\ref{eq:ktmax}) and (\ref{eq:Delta}) then imply that $R_{\mathrm{max}}$ depends mainly on the ratios $v_b/v_e$ and $v_s/v_e$, not on the individual speeds. As $v_b/v_e$ decreases, $k_{T\mathrm{max}}$ increases, and hence $R_{\mathrm{max}}$ increases. Since $v_s/v_e \ll 1$ unless $T_i \gg T_e$, the final term on the right hand side of Eq.~(\ref{eq:Delta}) can be neglected, so $R_{\mathrm{max}}$ decreases with increasing $v_s/v_e$. The dependence of $R_{\mathrm{max}}$ on $v_b/v_e$ is stronger than on $v_s/v_e$, which can be seen in Fig.~\ref{fig:rmax}.

The significance of the FOTC to the cold-plasma rate of second harmonic emission is now assessed for different applications. In coronal type III solar radio bursts, typical parameters are $v_b\approx (0.2-0.5)c$, $v_e\approx 0.02c$ and $v_s=1.5\times 10^{-4}c$. This gives $k_{T\mathrm{max}}/k_{T0}=1.02-1.003$ from Eqs.~(\ref{eq:ktmax}) and (\ref{eq:Delta}), and hence $R_{\mathrm{max}}=0.08-0.01$ from Eq.~(\ref{eq:R}). However, \citet{dulk87} determined much lower electron beam speeds from their observations, ranging from $v_b=(0.07-0.25)c$, with an average of $0.14c$. These values lead to the range $k_{T\mathrm{max}}/k_{T0}=1.16-1.04$ and $R_{\mathrm{max}}=0.66-0.05$, with an average of $k_{T\mathrm{max}}/k_{T0}=1.04$ and $R_{\mathrm{max}}=0.16$. Thus, the second harmonic emission rate may be well approximated by assuming a cold plasma for faster beams, but the FOTC becomes important for the slower electron beams measured by Dulk \textit{et al}.  

\begin{figure}
  \begin{center}
    \includegraphics{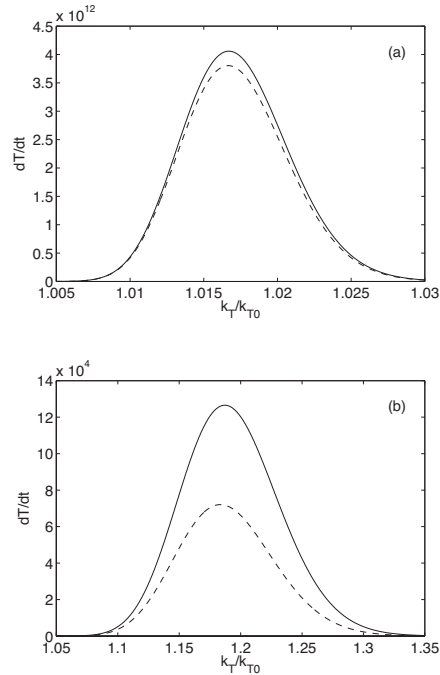}
  \end{center}
  \caption{Second harmonic emission rate for (a) coronal type III burst parameters: $v_e=0.02c$, $v_b/v_e = 10$, and $v_s/v_e=0.03$; and (b) Earth's foreshock parameters: $v_e=0.003c$, $v_b/v_e=3$, and $v_s/v_e=0.1$. Dashed lines are for a cold plasma while solid lines include the FOTC.}
  \label{fig:rate}
\end{figure}

The electron beams responsible for significant radio emission are typically much slower in the ``foreshock'' regions upstream of shocks. Examples are Earth's foreshock radio emissions, produced upstream of Earth's bow shock, and type II solar radio bursts associated with traveling shocks. At Earth's foreshock, $v_e\approx 3\times 10^{-3}c$, $v_s\approx 3\times 10^{-4}c$, and $v_b/v_e\approx 2-10$ (Ref.~\onlinecite{kuncic04}), which gives $k_{T\mathrm{max}}/k_{T0}=1.44-1.01$ and $R_{\mathrm{max}}=1.82-0.04$. \citet{knock01} calculated, for interplanetary type II bursts, a maximum in the emissivity of second harmonic radiation where $v_b/v_e\approx 3.5$ for a thermal speed $v_e=0.005c$. In this case, taking $v_s=1.3\times 10^{-4}c$ leads to $k_{T\mathrm{max}}/k_{T0}=1.15$ and $R_{\mathrm{max}}=0.63$. These values of $R_{\mathrm{max}}$ indicate that the FOTC may be a significant contribution to the total rate in foreshock emissions, and can even exceed the cold-plasma contribution. Figure \ref{fig:rate} shows the emission rate versus $k_T$ for typical coronal type III burst and Earth's foreshock parameters. Notably, the peak wavenumber $k_{T\mathrm{max}}$ stays almost constant when the FOTC is added to the emission rate.

Assuming the first term in $R(k_T)$ in Eq.~(\ref{eq:R}) to be negligible, and also that $v_s/v_e\ll 1$, we obtain $R>1$ for $v_b/v_e<2.9$. Thus, for sufficiently slow electron beams, the contribution from the FOTC exceeds the cold-plasma contribution to the emission rate. However, the assumption made in Eq.~(\ref{eq:omegaL}) that $k_L\ll \lambda_D^{-1}$, which corresponds approximately to $v_b/v_e\gg 1$, is not satisfied very well for these slow foreshock parameters. Thus for small $v_b/v_e$ the expression for $\cos\psi(k_1,k_T)$ given in Eq.~(\ref{eq:cospsi}), and hence the rate in Eq.~(\ref{eq:dTdt}), will be less accurate.

\section{Summary and conclusion}

Both the cold-plasma and the exact quadratic response tensor (QRT) describe three-wave interactions in which each wave is fast (that is, its phase speed is greater than the thermal speed). However, neither is ideal for the calculation of interaction rates: the cold-plasma QRT is readily calculable, but is only accurate where all phase speeds are much greater than the thermal speed; conversely, the exact QRT provides an accurate description of three-wave interactions between any waves, but its direct application to the calculation of rates is infeasible. The approximate QRT that we have derived here, which is the sum of the cold-plasma QRT and its first-order thermal correction (FOTC), overcomes these disadvantages since it is more accurate than the cold-plasma QRT alone but still permits a calculation of the interaction rate. It is also valid for arbitrary isotropic velocity distributions. This approximate QRT is therefore suitable for modeling three-wave interactions in space plasmas, in which thermal effects are important for the interacting waves, and the velocity distributions are commonly non-Maxwellian.

The rate of second harmonic plasma emission via Langmuir-wave coalescence has previously been treated with the cold-plasma QRT. Therefore, the resulting expression is inaccurate where the phase speed of one or more of the waves is similar to the thermal speed. Using our result for the approximate QRT, we have derived the FOTC to the rate of second harmonic plasma emission. The ratio of the FOTC rate to the cold-plasma rate is easily calculated using Eq.~(\ref{eq:R}); it is only a function of the transverse wave number $k_T$, and does not require an integral over the Langmuir wave number. The importance of the FOTC to the emission rate is determined by the ratios $v_b/v_e$ and $v_s/v_e$: the FOTC to the emission rate becomes larger compared to the cold-plasma emission rate as both $v_b/v_e$ and $v_s/v_e$ decrease. The FOTC to the cold-plasma emission rate is therefore important in foreshock emission, where the electron beam speed is not much larger than the electron thermal speed (within a factor of $\sim 2-10$). In the case where $v_s/v_e\ll 1$, the FOTC to the cold-plasma emission rate is greater than the cold-plasma emission rate for $v_b/v_e\lesssim 3$. 

Future work will involve deriving more accurate expressions for the rates of the processes $L\rightarrow L'+S$ and $L+S\rightarrow T$ involved in plasma emission.

\begin{acknowledgments}

The Australian Research Council and an Australian Postgraduate Award supported this work.

\end{acknowledgments}

\providecommand{\noopsort}[1]{}\providecommand{\singleletter}[1]{#1}%


\begin{thebibliography}{39}%
\makeatletter
\providecommand \@ifxundefined [1]{%
 \@ifx{#1\undefined}
}%
\providecommand \@ifnum [1]{%
 \ifnum #1\expandafter \@firstoftwo
 \else \expandafter \@secondoftwo
 \fi
}%
\providecommand \@ifx [1]{%
 \ifx #1\expandafter \@firstoftwo
 \else \expandafter \@secondoftwo
 \fi
}%
\providecommand \natexlab [1]{#1}%
\providecommand \enquote  [1]{``#1''}%
\providecommand \bibnamefont  [1]{#1}%
\providecommand \bibfnamefont [1]{#1}%
\providecommand \citenamefont [1]{#1}%
\providecommand \href@noop [0]{\@secondoftwo}%
\providecommand \href [0]{\begingroup \@sanitize@url \@href}%
\providecommand \@href[1]{\@@startlink{#1}\@@href}%
\providecommand \@@href[1]{\endgroup#1\@@endlink}%
\providecommand \@sanitize@url [0]{\catcode `\\12\catcode `\$12\catcode
  `\&12\catcode `\#12\catcode `\^12\catcode `\_12\catcode `\%12\relax}%
\providecommand \@@startlink[1]{}%
\providecommand \@@endlink[0]{}%
\providecommand \url  [0]{\begingroup\@sanitize@url \@url }%
\providecommand \@url [1]{\endgroup\@href {#1}{\urlprefix }}%
\providecommand \urlprefix  [0]{URL }%
\providecommand \Eprint [0]{\href }%
\@ifxundefined \urlstyle {%
  \providecommand \doi  [0]{\begingroup \@sanitize@url \@doi}%
  \providecommand \@doi [1]{\endgroup \@@startlink {\doibase
  #1}doi:\discretionary {}{}{}#1\@@endlink }%
}{%
  \providecommand \doi  [0]{doi:\discretionary{}{}{}\begingroup
  \urlstyle{rm}\Url }%
}%
\providecommand \doibase [0]{http://dx.doi.org/}%
\providecommand \Doi [0]{\begingroup \@sanitize@url \@Doi }%
\providecommand \@Doi  [1]{\endgroup\@@startlink{\doibase#1}\@@Doi}%
\providecommand \@@Doi [1]{#1\@@endlink}%
\providecommand \selectlanguage [0]{\@gobble}%
\providecommand \bibinfo  [0]{\@secondoftwo}%
\providecommand \bibfield  [0]{\@secondoftwo}%
\providecommand \translation [1]{[#1]}%
\providecommand \BibitemOpen [0]{}%
\providecommand \bibitemStop [0]{}%
\providecommand \bibitemNoStop [0]{.\EOS\space}%
\providecommand \EOS [0]{\spacefactor3000\relax}%
\providecommand \BibitemShut  [1]{\csname bibitem#1\endcsname}%
\bibitem [{\citenamefont {Wild}\ and\ \citenamefont {McCready}(1950)}]{wild50}%
  \BibitemOpen
  \bibfield  {author} {\bibinfo {author} {\bibfnamefont {J.~P.}\ \bibnamefont
  {Wild}}\ and\ \bibinfo {author} {\bibfnamefont {L.~L.}\ \bibnamefont
  {McCready}},\ }\href@noop {} {\bibfield  {journal} {\bibinfo  {journal}
  {Aust. J. Sci. Res. A},\ }\textbf {\bibinfo {volume} {3}},\ \bibinfo {pages}
  {387} (\bibinfo {year} {1950})}\BibitemShut {NoStop}%
\bibitem [{\citenamefont {Wild}\ \emph {et~al.}(1963)\citenamefont {Wild},
  \citenamefont {Smerd},\ and\ \citenamefont {Weiss}}]{wild63}%
  \BibitemOpen
  \bibfield  {author} {\bibinfo {author} {\bibfnamefont {J.~P.}\ \bibnamefont
  {Wild}}, \bibinfo {author} {\bibfnamefont {S.~F.}\ \bibnamefont {Smerd}}, \
  and\ \bibinfo {author} {\bibfnamefont {A.~A.}\ \bibnamefont {Weiss}},\
  }\href@noop {} {\bibfield  {journal} {\bibinfo  {journal} {Ann. Rev. Astron.
  Astrophys.},\ }\textbf {\bibinfo {volume} {1}},\ \bibinfo {pages} {291}
  (\bibinfo {year} {1963})}\BibitemShut {NoStop}%
\bibitem [{\citenamefont {Cane}\ \emph {et~al.}(1982)\citenamefont {Cane},
  \citenamefont {Stone}, \citenamefont {Fainberg}, \citenamefont {Steinberg},\
  and\ \citenamefont {Hoang}}]{cane82}%
  \BibitemOpen
  \bibfield  {author} {\bibinfo {author} {\bibfnamefont {H.~V.}\ \bibnamefont
  {Cane}}, \bibinfo {author} {\bibfnamefont {R.~G.}\ \bibnamefont {Stone}},
  \bibinfo {author} {\bibfnamefont {J.}~\bibnamefont {Fainberg}}, \bibinfo
  {author} {\bibfnamefont {J.~L.}\ \bibnamefont {Steinberg}}, \ and\ \bibinfo
  {author} {\bibfnamefont {S.}~\bibnamefont {Hoang}},\ }\href@noop {}
  {\bibfield  {journal} {\bibinfo  {journal} {Sol. Phys.},\ }\textbf {\bibinfo
  {volume} {78}},\ \bibinfo {pages} {187} (\bibinfo {year} {1982})}\BibitemShut
  {NoStop}%
\bibitem [{\citenamefont {Suzuki}\ and\ \citenamefont {Dulk}(1985)}]{suzuki85}%
  \BibitemOpen
  \bibfield  {author} {\bibinfo {author} {\bibfnamefont {S.}~\bibnamefont
  {Suzuki}}\ and\ \bibinfo {author} {\bibfnamefont {G.~A.}\ \bibnamefont
  {Dulk}},\ }in\ \href@noop {} {\emph {\bibinfo {booktitle} {Solar
  Radiophysics}}},\ \bibinfo {editor} {edited by\ \bibinfo {editor}
  {\bibfnamefont {D.~J.}\ \bibnamefont {McLean}}\ and\ \bibinfo {editor}
  {\bibfnamefont {N.~R.}\ \bibnamefont {Labrum}}}\ (\bibinfo  {publisher}
  {Cambridge University Press},\ \bibinfo {address} {Cambridge},\ \bibinfo
  {year} {1985})\ p.\ \bibinfo {pages} {289}\BibitemShut {NoStop}%
\bibitem [{\citenamefont {Robinson}\ and\ \citenamefont
  {Cairns}(2000)}]{robinson00}%
  \BibitemOpen
  \bibfield  {author} {\bibinfo {author} {\bibfnamefont {P.~A.}\ \bibnamefont
  {Robinson}}\ and\ \bibinfo {author} {\bibfnamefont {I.~H.}\ \bibnamefont
  {Cairns}},\ }in\ \href@noop {} {\emph {\bibinfo {booktitle} {Radio Astronomy
  at Long Wavelengths}}},\ \bibinfo {series} {Geophys. Monograph Ser.}, Vol.\
  \bibinfo {volume} {119},\ \bibinfo {editor} {edited by\ \bibinfo {editor}
  {\bibfnamefont {R.~G.}\ \bibnamefont {Stone}}, \bibinfo {editor}
  {\bibfnamefont {K.~W.}\ \bibnamefont {Weiler}}, \bibinfo {editor}
  {\bibfnamefont {M.~L.}\ \bibnamefont {Goldstein}}, \ and\ \bibinfo {editor}
  {\bibfnamefont {J.~L.}\ \bibnamefont {Bougeret}}}\ (\bibinfo  {publisher}
  {American Geophysical Union},\ \bibinfo {address} {Washington},\ \bibinfo
  {year} {2000})\ p.~\bibinfo {pages} {37}\BibitemShut {NoStop}%
\bibitem [{\citenamefont {Gurnett}(1975)}]{gurnett75}%
  \BibitemOpen
  \bibfield  {author} {\bibinfo {author} {\bibfnamefont {D.~A.}\ \bibnamefont
  {Gurnett}},\ }\href@noop {} {\bibfield  {journal} {\bibinfo  {journal} {J.
  Geophys. Res.},\ }\textbf {\bibinfo {volume} {80}},\ \bibinfo {pages} {2751}
  (\bibinfo {year} {1975})}\BibitemShut {NoStop}%
\bibitem [{\citenamefont {Hoang}\ \emph {et~al.}(1981)\citenamefont {Hoang},
  \citenamefont {Fainberg}, \citenamefont {Steinberg}, \citenamefont {Stone},\
  and\ \citenamefont {Zwickl}}]{hoang81}%
  \BibitemOpen
  \bibfield  {author} {\bibinfo {author} {\bibfnamefont {S.}~\bibnamefont
  {Hoang}}, \bibinfo {author} {\bibfnamefont {J.}~\bibnamefont {Fainberg}},
  \bibinfo {author} {\bibfnamefont {J.~L.}\ \bibnamefont {Steinberg}}, \bibinfo
  {author} {\bibfnamefont {R.~G.}\ \bibnamefont {Stone}}, \ and\ \bibinfo
  {author} {\bibfnamefont {R.~H.}\ \bibnamefont {Zwickl}},\ }\href@noop {}
  {\bibfield  {journal} {\bibinfo  {journal} {J. Geophys. Res.},\ }\textbf
  {\bibinfo {volume} {86}},\ \bibinfo {pages} {4531} (\bibinfo {year}
  {1981})}\BibitemShut {NoStop}%
\bibitem [{\citenamefont {Field}(1956)}]{field56}%
  \BibitemOpen
  \bibfield  {author} {\bibinfo {author} {\bibfnamefont {G.~B.}\ \bibnamefont
  {Field}},\ }\href@noop {} {\bibfield  {journal} {\bibinfo  {journal}
  {Astrophys. J.},\ }\textbf {\bibinfo {volume} {124}},\ \bibinfo {pages} {555}
  (\bibinfo {year} {1956})}\BibitemShut {NoStop}%
\bibitem [{\citenamefont {Melrose}(1980){\natexlab{a}}}]{melrose80a}%
  \BibitemOpen
  \bibfield  {author} {\bibinfo {author} {\bibfnamefont {D.~B.}\ \bibnamefont
  {Melrose}},\ }\href@noop {} {\bibfield  {journal} {\bibinfo  {journal} {Aust.
  J. Phys.},\ }\textbf {\bibinfo {volume} {33}},\ \bibinfo {pages} {121}
  (\bibinfo {year} {1980}{\natexlab{a}})}\BibitemShut {NoStop}%
\bibitem [{\citenamefont {Yin}\ \emph {et~al.}(1998)\citenamefont {Yin},
  \citenamefont {Ashour-Abdalla}, \citenamefont {El-Alaoui}, \citenamefont
  {Bosqued},\ and\ \citenamefont {Bougeret}}]{yin98}%
  \BibitemOpen
  \bibfield  {author} {\bibinfo {author} {\bibfnamefont {L.}~\bibnamefont
  {Yin}}, \bibinfo {author} {\bibfnamefont {M.}~\bibnamefont {Ashour-Abdalla}},
  \bibinfo {author} {\bibfnamefont {M.}~\bibnamefont {El-Alaoui}}, \bibinfo
  {author} {\bibfnamefont {J.~M.}\ \bibnamefont {Bosqued}}, \ and\ \bibinfo
  {author} {\bibfnamefont {J.~L.}\ \bibnamefont {Bougeret}},\ }\href@noop {}
  {\bibfield  {journal} {\bibinfo  {journal} {Geophys. Res. Lett.},\ }\textbf
  {\bibinfo {volume} {25}},\ \bibinfo {pages} {2609} (\bibinfo {year}
  {1998})}\BibitemShut {NoStop}%
\bibitem [{\citenamefont {Wu}\ \emph {et~al.}(2002)\citenamefont {Wu},
  \citenamefont {Wang}, \citenamefont {Yoon}, \citenamefont {Zheng},\ and\
  \citenamefont {Wang}}]{wu02}%
  \BibitemOpen
  \bibfield  {author} {\bibinfo {author} {\bibfnamefont {C.~S.}\ \bibnamefont
  {Wu}}, \bibinfo {author} {\bibfnamefont {C.~B.}\ \bibnamefont {Wang}},
  \bibinfo {author} {\bibfnamefont {P.~H.}\ \bibnamefont {Yoon}}, \bibinfo
  {author} {\bibfnamefont {H.~N.}\ \bibnamefont {Zheng}}, \ and\ \bibinfo
  {author} {\bibfnamefont {S.}~\bibnamefont {Wang}},\ }\href@noop {} {\bibfield
   {journal} {\bibinfo  {journal} {Astrophys. J.},\ }\textbf {\bibinfo {volume}
  {575}},\ \bibinfo {pages} {1094} (\bibinfo {year} {2002})}\BibitemShut
  {NoStop}%
\bibitem [{\citenamefont {Ginzburg}\ and\ \citenamefont
  {Zheleznyakov}(1958)}]{ginzburg58}%
  \BibitemOpen
  \bibfield  {author} {\bibinfo {author} {\bibfnamefont {V.~L.}\ \bibnamefont
  {Ginzburg}}\ and\ \bibinfo {author} {\bibfnamefont {V.~V.}\ \bibnamefont
  {Zheleznyakov}},\ }\href@noop {} {\bibfield  {journal} {\bibinfo  {journal}
  {Sov. Astron.},\ }\textbf {\bibinfo {volume} {2}},\ \bibinfo {pages} {653}
  (\bibinfo {year} {1958})}\BibitemShut {NoStop}%
\bibitem [{\citenamefont {Melrose}(1980){\natexlab{b}}}]{melrose80c}%
  \BibitemOpen
  \bibfield  {author} {\bibinfo {author} {\bibfnamefont {D.~B.}\ \bibnamefont
  {Melrose}},\ }\href@noop {} {\bibfield  {journal} {\bibinfo  {journal} {Space
  Sci. Rev.},\ }\textbf {\bibinfo {volume} {26}},\ \bibinfo {pages} {3}
  (\bibinfo {year} {1980}{\natexlab{b}})}\BibitemShut {NoStop}%
\bibitem [{\citenamefont {Nelson}\ and\ \citenamefont
  {Melrose}(1985)}]{melrose85}%
  \BibitemOpen
  \bibfield  {author} {\bibinfo {author} {\bibfnamefont {G.~J.}\ \bibnamefont
  {Nelson}}\ and\ \bibinfo {author} {\bibfnamefont {D.~B.}\ \bibnamefont
  {Melrose}},\ }in\ \href@noop {} {\emph {\bibinfo {booktitle} {Solar
  Radiophysics}}},\ \bibinfo {editor} {edited by\ \bibinfo {editor}
  {\bibfnamefont {D.~J.}\ \bibnamefont {McLean}}\ and\ \bibinfo {editor}
  {\bibfnamefont {N.~R.}\ \bibnamefont {Labrum}}}\ (\bibinfo  {publisher}
  {Cambridge University Press},\ \bibinfo {address} {Cambridge},\ \bibinfo
  {year} {1985})\ p.\ \bibinfo {pages} {333}\BibitemShut {NoStop}%
\bibitem [{\citenamefont {Cairns}\ and\ \citenamefont
  {Melrose}(1985)}]{cairns85}%
  \BibitemOpen
  \bibfield  {author} {\bibinfo {author} {\bibfnamefont {I.~H.}\ \bibnamefont
  {Cairns}}\ and\ \bibinfo {author} {\bibfnamefont {D.~B.}\ \bibnamefont
  {Melrose}},\ }\href@noop {} {\bibfield  {journal} {\bibinfo  {journal} {J.
  Geophys. Res.},\ }\textbf {\bibinfo {volume} {90}},\ \bibinfo {pages} {6637}
  (\bibinfo {year} {1985})}\BibitemShut {NoStop}%
\bibitem [{\citenamefont {Tsytovich}(1970)}]{tsytovich70}%
  \BibitemOpen
  \bibfield  {author} {\bibinfo {author} {\bibfnamefont {V.~N.}\ \bibnamefont
  {Tsytovich}},\ }\href@noop {} {\emph {\bibinfo {title} {Nonlinear Effects in
  a Plasma}}}\ (\bibinfo  {publisher} {Plenum},\ \bibinfo {address} {New
  York},\ \bibinfo {year} {1970})\BibitemShut {NoStop}%
\bibitem [{\citenamefont {Sitenko}(1982)}]{sitenko82}%
  \BibitemOpen
  \bibfield  {author} {\bibinfo {author} {\bibfnamefont {A.~G.}\ \bibnamefont
  {Sitenko}},\ }\href@noop {} {\emph {\bibinfo {title} {Fluctuations and
  Non-linear Wave Interactions in Plasmas}}}\ (\bibinfo  {publisher}
  {Pergamon},\ \bibinfo {address} {Oxford},\ \bibinfo {year}
  {1982})\BibitemShut {NoStop}%
\bibitem [{\citenamefont {Melrose}(1980){\natexlab{c}}}]{melrose80b}%
  \BibitemOpen
  \bibfield  {author} {\bibinfo {author} {\bibfnamefont {D.~B.}\ \bibnamefont
  {Melrose}},\ }\href@noop {} {\emph {\bibinfo {title} {Plasma Astrophysics
  Volume {II}}}}\ (\bibinfo  {publisher} {Gordon \& Breach},\ \bibinfo
  {address} {New York},\ \bibinfo {year} {1980})\BibitemShut {NoStop}%
\bibitem [{\citenamefont {Sagdeev}\ and\ \citenamefont
  {Galeev}(1969)}]{sagdeev69}%
  \BibitemOpen
  \bibfield  {author} {\bibinfo {author} {\bibfnamefont {R.~Z.}\ \bibnamefont
  {Sagdeev}}\ and\ \bibinfo {author} {\bibfnamefont {A.~A.}\ \bibnamefont
  {Galeev}},\ }\href@noop {} {\emph {\bibinfo {title} {Nonlinear plasma
  theory}}}\ (\bibinfo  {publisher} {Benjamin},\ \bibinfo {address} {New
  York},\ \bibinfo {year} {1969})\BibitemShut {NoStop}%
\bibitem [{\citenamefont {Tsytovich}(1977)}]{tsytovich77}%
  \BibitemOpen
  \bibfield  {author} {\bibinfo {author} {\bibfnamefont {V.~N.}\ \bibnamefont
  {Tsytovich}},\ }\href@noop {} {\emph {\bibinfo {title} {An Introduction to
  the Theory of Plasma Turbulence}}}\ (\bibinfo  {publisher} {Pergamon},\
  \bibinfo {address} {New York},\ \bibinfo {year} {1977})\BibitemShut {NoStop}%
\bibitem [{\citenamefont {Percival}\ and\ \citenamefont
  {Robinson}(1998){\natexlab{a}}}]{percival98a}%
  \BibitemOpen
  \bibfield  {author} {\bibinfo {author} {\bibfnamefont {D.~J.}\ \bibnamefont
  {Percival}}\ and\ \bibinfo {author} {\bibfnamefont {P.~A.}\ \bibnamefont
  {Robinson}},\ }\href@noop {} {\bibfield  {journal} {\bibinfo  {journal}
  {Phys. Plasmas},\ }\textbf {\bibinfo {volume} {5}},\ \bibinfo {pages} {1279}
  (\bibinfo {year} {1998}{\natexlab{a}})}\BibitemShut {NoStop}%
\bibitem [{\citenamefont {Percival}\ and\ \citenamefont
  {Robinson}(1998){\natexlab{b}}}]{percival98b}%
  \BibitemOpen
  \bibfield  {author} {\bibinfo {author} {\bibfnamefont {D.~J.}\ \bibnamefont
  {Percival}}\ and\ \bibinfo {author} {\bibfnamefont {P.~A.}\ \bibnamefont
  {Robinson}},\ }\href@noop {} {\bibfield  {journal} {\bibinfo  {journal} {J.
  Math. Phys.},\ }\textbf {\bibinfo {volume} {39}},\ \bibinfo {pages} {3678}
  (\bibinfo {year} {1998}{\natexlab{b}})}\BibitemShut {NoStop}%
\bibitem [{\citenamefont {Percival}(1992)}]{percival_thesis}%
  \BibitemOpen
  \bibfield  {author} {\bibinfo {author} {\bibfnamefont {D.~J.}\ \bibnamefont
  {Percival}},\ }\emph {\bibinfo {title} {Nonlinear wave processes in
  collisionless plasmas}},\ \href@noop {} {Ph.D. thesis},\ \bibinfo  {school}
  {University of Sydney} (\bibinfo {year} {1992})\BibitemShut {NoStop}%
\bibitem [{\citenamefont {Vasyliunas}(1968)}]{vasyliunas68}%
  \BibitemOpen
  \bibfield  {author} {\bibinfo {author} {\bibfnamefont {V.~M.}\ \bibnamefont
  {Vasyliunas}},\ }\href@noop {} {\bibfield  {journal} {\bibinfo  {journal} {J.
  Geophys. Res.},\ }\textbf {\bibinfo {volume} {73}},\ \bibinfo {pages} {2839}
  (\bibinfo {year} {1968})}\BibitemShut {NoStop}%
\bibitem [{\citenamefont {Maksimovic}\ \emph {et~al.}(1997)\citenamefont
  {Maksimovic}, \citenamefont {Pierrard},\ and\ \citenamefont
  {Riley}}]{maksimovic97}%
  \BibitemOpen
  \bibfield  {author} {\bibinfo {author} {\bibfnamefont {M.}~\bibnamefont
  {Maksimovic}}, \bibinfo {author} {\bibfnamefont {V.}~\bibnamefont
  {Pierrard}}, \ and\ \bibinfo {author} {\bibfnamefont {P.}~\bibnamefont
  {Riley}},\ }\href@noop {} {\bibfield  {journal} {\bibinfo  {journal}
  {Geophys. Res. Lett.},\ }\textbf {\bibinfo {volume} {24}},\ \bibinfo {pages}
  {1151} (\bibinfo {year} {1997})}\BibitemShut {NoStop}%
\bibitem [{\citenamefont {Cairns}(1987){\natexlab{a}}}]{cairns87a}%
  \BibitemOpen
  \bibfield  {author} {\bibinfo {author} {\bibfnamefont {I.~H.}\ \bibnamefont
  {Cairns}},\ }\href@noop {} {\bibfield  {journal} {\bibinfo  {journal} {J.
  Plasma Physics},\ }\textbf {\bibinfo {volume} {38}},\ \bibinfo {pages} {179}
  (\bibinfo {year} {1987}{\natexlab{a}})}\BibitemShut {NoStop}%
\bibitem [{\citenamefont {Willes}\ \emph {et~al.}(1996)\citenamefont {Willes},
  \citenamefont {Robinson},\ and\ \citenamefont {Melrose}}]{willes96}%
  \BibitemOpen
  \bibfield  {author} {\bibinfo {author} {\bibfnamefont {A.~J.}\ \bibnamefont
  {Willes}}, \bibinfo {author} {\bibfnamefont {P.~A.}\ \bibnamefont
  {Robinson}}, \ and\ \bibinfo {author} {\bibfnamefont {D.~B.}\ \bibnamefont
  {Melrose}},\ }\href@noop {} {\bibfield  {journal} {\bibinfo  {journal} {Phys.
  Plasmas},\ }\textbf {\bibinfo {volume} {3}},\ \bibinfo {pages} {149}
  (\bibinfo {year} {1996})}\BibitemShut {NoStop}%
\bibitem [{\citenamefont {Li}\ \emph {et~al.}(2005)\citenamefont {Li},
  \citenamefont {Willes}, \citenamefont {Robinson},\ and\ \citenamefont
  {Cairns}}]{li05}%
  \BibitemOpen
  \bibfield  {author} {\bibinfo {author} {\bibfnamefont {B.}~\bibnamefont
  {Li}}, \bibinfo {author} {\bibfnamefont {A.~J.}\ \bibnamefont {Willes}},
  \bibinfo {author} {\bibfnamefont {P.~A.}\ \bibnamefont {Robinson}}, \ and\
  \bibinfo {author} {\bibfnamefont {I.~H.}\ \bibnamefont {Cairns}},\
  }\href@noop {} {\bibfield  {journal} {\bibinfo  {journal} {Phys. Plasmas},\
  }\textbf {\bibinfo {volume} {12}},\ \bibinfo {pages} {012103} (\bibinfo
  {year} {2005})}\BibitemShut {NoStop}%
\bibitem [{\citenamefont {Zlotnik}(1978)}]{zlotnik78}%
  \BibitemOpen
  \bibfield  {author} {\bibinfo {author} {\bibfnamefont {E.~Y.}\ \bibnamefont
  {Zlotnik}},\ }\href@noop {} {\bibfield  {journal} {\bibinfo  {journal} {Sov.
  Astron.},\ }\textbf {\bibinfo {volume} {22}},\ \bibinfo {pages} {228}
  (\bibinfo {year} {1978})}\BibitemShut {NoStop}%
\bibitem [{\citenamefont {Cairns}(1987){\natexlab{b}}}]{cairns87b}%
  \BibitemOpen
  \bibfield  {author} {\bibinfo {author} {\bibfnamefont {I.~H.}\ \bibnamefont
  {Cairns}},\ }\href@noop {} {\bibfield  {journal} {\bibinfo  {journal} {J.
  Plasma Physics},\ }\textbf {\bibinfo {volume} {38}},\ \bibinfo {pages} {199}
  (\bibinfo {year} {1987}{\natexlab{b}})}\BibitemShut {NoStop}%
\bibitem [{\citenamefont {Rhee}\ \emph {et~al.}(2009)\citenamefont {Rhee},
  \citenamefont {Ryu}, \citenamefont {Woo}, \citenamefont {Kaang},
  \citenamefont {Yi},\ and\ \citenamefont {Yoon}}]{rhee09}%
  \BibitemOpen
  \bibfield  {author} {\bibinfo {author} {\bibfnamefont {T.}~\bibnamefont
  {Rhee}}, \bibinfo {author} {\bibfnamefont {C.~M.}\ \bibnamefont {Ryu}},
  \bibinfo {author} {\bibfnamefont {M.}~\bibnamefont {Woo}}, \bibinfo {author}
  {\bibfnamefont {H.~H.}\ \bibnamefont {Kaang}}, \bibinfo {author}
  {\bibfnamefont {S.}~\bibnamefont {Yi}}, \ and\ \bibinfo {author}
  {\bibfnamefont {P.~H.}\ \bibnamefont {Yoon}},\ }\href@noop {} {\bibfield
  {journal} {\bibinfo  {journal} {Astrophys. J.},\ }\textbf {\bibinfo {volume}
  {694}},\ \bibinfo {pages} {618} (\bibinfo {year} {2009})}\BibitemShut
  {NoStop}%
\bibitem [{\citenamefont {Melrose}(1982)}]{melrose82}%
  \BibitemOpen
  \bibfield  {author} {\bibinfo {author} {\bibfnamefont {D.~B.}\ \bibnamefont
  {Melrose}},\ }\href@noop {} {\bibfield  {journal} {\bibinfo  {journal} {Sol.
  Phys.},\ }\textbf {\bibinfo {volume} {79}},\ \bibinfo {pages} {173} (\bibinfo
  {year} {1982})}\BibitemShut {NoStop}%
\bibitem [{\citenamefont {Robinson}\ and\ \citenamefont
  {Cairns}(1998){\natexlab{a}}}]{robinson98a}%
  \BibitemOpen
  \bibfield  {author} {\bibinfo {author} {\bibfnamefont {P.~A.}\ \bibnamefont
  {Robinson}}\ and\ \bibinfo {author} {\bibfnamefont {I.~H.}\ \bibnamefont
  {Cairns}},\ }\href@noop {} {\bibfield  {journal} {\bibinfo  {journal} {Sol.
  Phys.},\ }\textbf {\bibinfo {volume} {181}},\ \bibinfo {pages} {363}
  (\bibinfo {year} {1998}{\natexlab{a}})}\BibitemShut {NoStop}%
\bibitem [{\citenamefont {Robinson}\ and\ \citenamefont
  {Cairns}(1998){\natexlab{b}}}]{robinson98b}%
  \BibitemOpen
  \bibfield  {author} {\bibinfo {author} {\bibfnamefont {P.~A.}\ \bibnamefont
  {Robinson}}\ and\ \bibinfo {author} {\bibfnamefont {I.~H.}\ \bibnamefont
  {Cairns}},\ }\href@noop {} {\bibfield  {journal} {\bibinfo  {journal} {Sol.
  Phys.},\ }\textbf {\bibinfo {volume} {181}},\ \bibinfo {pages} {395}
  (\bibinfo {year} {1998}{\natexlab{b}})}\BibitemShut {NoStop}%
\bibitem [{\citenamefont {Robinson}\ and\ \citenamefont
  {Cairns}(1998){\natexlab{c}}}]{robinson98c}%
  \BibitemOpen
  \bibfield  {author} {\bibinfo {author} {\bibfnamefont {P.~A.}\ \bibnamefont
  {Robinson}}\ and\ \bibinfo {author} {\bibfnamefont {I.~H.}\ \bibnamefont
  {Cairns}},\ }\href@noop {} {\bibfield  {journal} {\bibinfo  {journal} {Sol.
  Phys.},\ }\textbf {\bibinfo {volume} {181}},\ \bibinfo {pages} {429}
  (\bibinfo {year} {1998}{\natexlab{c}})}\BibitemShut {NoStop}%
\bibitem [{\citenamefont {Thorne}\ and\ \citenamefont
  {Summers}(1991)}]{thorne91}%
  \BibitemOpen
  \bibfield  {author} {\bibinfo {author} {\bibfnamefont {R.~M.}\ \bibnamefont
  {Thorne}}\ and\ \bibinfo {author} {\bibfnamefont {D.}~\bibnamefont
  {Summers}},\ }\href@noop {} {\bibfield  {journal} {\bibinfo  {journal} {Phys.
  Fluids B},\ }\textbf {\bibinfo {volume} {3}},\ \bibinfo {pages} {2117}
  (\bibinfo {year} {1991})}\BibitemShut {NoStop}%
\bibitem [{\citenamefont {Dulk}\ \emph {et~al.}(1987)\citenamefont {Dulk},
  \citenamefont {Steinberg}, \citenamefont {Hoang},\ and\ \citenamefont
  {Goldman}}]{dulk87}%
  \BibitemOpen
  \bibfield  {author} {\bibinfo {author} {\bibfnamefont {G.~A.}\ \bibnamefont
  {Dulk}}, \bibinfo {author} {\bibfnamefont {J.~L.}\ \bibnamefont {Steinberg}},
  \bibinfo {author} {\bibfnamefont {S.}~\bibnamefont {Hoang}}, \ and\ \bibinfo
  {author} {\bibfnamefont {M.~V.}\ \bibnamefont {Goldman}},\ }\href@noop {}
  {\bibfield  {journal} {\bibinfo  {journal} {Astron. Astrophys.},\ }\textbf
  {\bibinfo {volume} {173}},\ \bibinfo {pages} {366} (\bibinfo {year}
  {1987})}\BibitemShut {NoStop}%
\bibitem [{\citenamefont {Kuncic}\ \emph {et~al.}(2004)\citenamefont {Kuncic},
  \citenamefont {Cairns},\ and\ \citenamefont {Knock}}]{kuncic04}%
  \BibitemOpen
  \bibfield  {author} {\bibinfo {author} {\bibfnamefont {Z.}~\bibnamefont
  {Kuncic}}, \bibinfo {author} {\bibfnamefont {I.~H.}\ \bibnamefont {Cairns}},
  \ and\ \bibinfo {author} {\bibfnamefont {S.~A.}\ \bibnamefont {Knock}},\
  }\href@noop {} {\bibfield  {journal} {\bibinfo  {journal} {J. Geophys.
  Res.},\ }\textbf {\bibinfo {volume} {109}},\ \bibinfo {pages} {2108}
  (\bibinfo {year} {2004})}\BibitemShut {NoStop}%
\bibitem [{\citenamefont {Knock}\ \emph {et~al.}(2001)\citenamefont {Knock},
  \citenamefont {Cairns}, \citenamefont {Robinson},\ and\ \citenamefont
  {Kuncic}}]{knock01}%
  \BibitemOpen
  \bibfield  {author} {\bibinfo {author} {\bibfnamefont {S.~A.}\ \bibnamefont
  {Knock}}, \bibinfo {author} {\bibfnamefont {I.~H.}\ \bibnamefont {Cairns}},
  \bibinfo {author} {\bibfnamefont {P.~A.}\ \bibnamefont {Robinson}}, \ and\
  \bibinfo {author} {\bibfnamefont {Z.}~\bibnamefont {Kuncic}},\ }\href@noop {}
  {\bibfield  {journal} {\bibinfo  {journal} {J. Geophys. Res.},\ }\textbf
  {\bibinfo {volume} {106}},\ \bibinfo {pages} {25041} (\bibinfo {year}
  {2001})}\BibitemShut {NoStop}%
\end{thebibliography}
\end{document}